\documentclass{PoS}

\usepackage{amsmath}
\usepackage{bm}
\usepackage{bbm}
\usepackage{subfigure}

\newcommand{\cN}{{\cal N}}

\newcommand{\cP}{{\cal P}}

\newcommand{\e}{\text{e}}
\newcommand{\ii}{\mathrm{i}}
\DeclareMathOperator{\Det}{Det}

\DeclareMathOperator{\tr}{tr}

\newcommand{\rk}{\right)}
\newcommand{\lk}{\left(}
\newcommand{\vsigma}{{\vec{\sigma}}}
\newcommand{\bra}[1]{\langle #1\rvert}
\newcommand{\ket}[1]{\lvert#1\rangle}

\newcommand{\be}{\begin{equation}}
\newcommand{\ee}{\end{equation}}
\newcommand{\sli}{\sum\limits}
\newcommand{\bal}{\begin{align}}
\newcommand{\eal}{\end{align}}
\newcommand{\bea}{\begin{eqnarray}}
\newcommand{\eea}{\end{eqnarray}}

\newcommand*{\vev}[1]{\left< #1 \right>}
\providecommand*{\coloneq}{\mathrel{\mathop:}=}
\newcommand*{\dd}[1][]{\mathop{\mathrm{d}^{#1}}\mkern-4mu}

\newcommand{\vA}{\vec{A}}
\newcommand{\vx}{\vec{x}}
\newcommand{\vp}{\vec{p}}
\newcommand{\va}{\vec{a}}
\newcommand{\vd}{\vec{d}}

\newcommand{\vD}{\vec{D}}

\newcommand{\ve}{\vec{e}}

\newcommand{\bR}{\mathbbm{R}}

\title{The effective potential of the confinement order parameter in the Hamiltonian Approach}

\ShortTitle{The effective potential of the confinement order parameter in the Hamiltonian Approach}

\author{\speaker{Hugo Reinhardt}\\
Universit\"at T\"ubingen, Institut f\"ur Theoretische Physik\\
Auf der Morgenstelle 14, 72076 T\"ubingen, Germany\\
E-mail: \email{hugo.reinhardt@uni-tuebingen.de}}

\author{Jan Heffner\\
Universit\"at T\"ubingen, Institut f\"ur Theoretische Physik\\
Auf der Morgenstelle 14, 72076 T\"ubingen, Germany
}
\abstract{
The effective potential of the order parameter for confinement is calculated within 
the variational approach to the Hamilton formulation of
 Yang--Mills theory. Compactifying one spatial dimension and using a background gauge fixing
this potential is obtained by minimizing the energy density for a given constant and color diagonal 
background field directed along the compactified dimension. 
Using Gaussian type trial wave functionals I establish an analytic relation between the propagators in the
background gauge at finite temperature and the corresponding zero temperature propagators in 
Coulomb gauge. In the simplest truncation, neglecting the ghost and using the ultraviolet form of the 
gluon energy one recovers the Weiss potential. On the other hand from the infrared form 
of the gluon energy one finds an effective potential which yields a vanishing Polyakov 
loop indicating the confined phase. From the full non-perturbative potential (with the ghost
included) one extracts a critical temperature of the deconfinement phase transition of 269
 MeV for the 
gauge group SU$(2)$ and 283 MeV for SU$(3)$.}

\FullConference{QCD-TNT-III-From quarks and gluons to hadronic matter: A bridge too far?\\
		2-6 September, 2013\\
		European Centre for Theoretical Studies in Nuclear Physics and Related Areas (ECT*), Villazzano, Trento (Italy)}
\begin{document}
\section{Introduction}\label{sectionI}
Understanding the deconfinement phase transition is one of the major challenges of particle physics. In quenched QCD reliable results are obtained within the lattice approach. This approach fails, however, at large baryon density due to the notorious fermion sign problem. Therefore alternative non-perturbative approaches to continuum QCD are desirable. In recent years a variational approach to Yang--Mills theory in Coulomb gauge was developed \cite{FeuRei04}, which has provided a decent description of the infrared sector of the theory \cite{EppReiSch07, SchLedRei06, Campagnari:2008yg,ReiEpp07,Pak:2009em,Reinhardt:2008ek}. Recently this approach was extended to finite temperature \cite{HefReiCam12} and also to full QCD \cite{Pak:2011wu,Pak:2013uba}. In this talk I will report on the calculation of the effective potential of the confinement 
order parameter within the Hamiltonian approach \cite{Reinhardt:2012qe, Reinhardt:2013iia}.

In quantum field theory the temperature $T$ is most easily introduced by compactifying the Euclidean time and interpreting the length $L$
of the compactified time interval as inverse temperature. In finite temperature SU($N$) Yang--Mills theory the order parameter of confinement
is the expectation value of the Polyakov loop
\be
\label{1}
P [A_0] = \frac{1}{N} \tr \cP \exp \left[ - \int^L_0 \dd x_0 A_0 \lk x_0, \vx \rk \right] \, .
\ee
The quantity $\langle P [A_0] (\vx) \rangle \sim \exp \left[ - F_\infty (\vx) L \right]$ is related to the free energy $F_\infty (\vx)$ of a (infinitely heavy)
quark at spatial position $\vx$. In the confined phase this quantity vanishes by center symmetry while it is non-zero in the deconfined
phase, where center symmetry is broken. In continuum Yang--Mills theory the Polyakov loop is most easily calculated in Polyakov gauge 
$\partial_0 A_0 = 0 \, , \, A_0 =$ diagonal. In the fundamental modular region $0 < A_0 L / 2 < \pi$ the Polyakov loop
$P [A_0]$ is (at least for the gauge groups SU$(2)$ and SU$(3)$) 
a unique function of the field $A_0$, which, for SU$(2)$, is given by $P [A_0] = \cos \lk A_0 L / 2 \rk$. As a consequence 
of this relation and of Jenssen's inequality instead of $\langle P [A_0] \rangle$ one can use alternatively $P  [\langle A_0 \rangle]$ or
$\langle A_0 \rangle$ as order parameter of confinement, see refs.~\cite{Marhauser:2008fz, Braun:2007bx}. 
Thus the order parameter of confinement can be most easily obtained by calculating the effective potential $e [a_0]$ of a 
temporal background field $a_0$ chosen in the Polyakov gauge and by calculating the Polyakov line (\ref{1}) from 
the field configuration $a^\text{min}_0$ which minimizes $e [a_0]$, i.e. $\langle P [A_0] \rangle \simeq P [a^\text{min}_0]$. The
effective potential $e [a_0]$ was first calculated in refs.~\cite{Gross:1980br, Weiss:1980rj} in 1-loop perturbation theory and is shown in 
fig.~\ref{fig1-1}. This potential is minimal for a vanishing field and the order parameter accordingly yields $P \left[a^\text{min}_0= 0 \right] = 1$, which indicates
the deconfining phase. In this talk I report on a non-perturbative evaluation of $e [a_0]$ \cite{Reinhardt:2012qe, Reinhardt:2013iia} in the Hamilton approach to Yang--Mills
theory \cite{FeuRei04}. For recent alternative work on the Polyakov loop see refs.~\cite{Fister:2013bh,Haas:2013qwp,Fischer:2013eca,Smith:2013msa,Langelage:2010yr,Diakonov:2012dx,Greensite:2012dy,Greensite:2013yd}.

It is obvious that the effective potential of $\langle A_0 \rangle$ cannot be straightforwardly evaluated in the Hamiltonian approach since the
letter assumes Weyl gauge $A_0 = 0$. However, we can exploit O$(4)$ invariance of Euclidean quantum field theory and compactify instead of the
time one spatial axis (for example the $x_3$-axis) to a circle and interpret the length $L$ of the compactified dimension as 
inverse temperature. (For more details see refs.~\cite{Reinhardt:2012qe, Reinhardt:2013iia}.)
Therefore we will consider in the following Yang--Mills theory at a finite compactified length $L$ in a constant color diagonal background field
$\va$ and calculate the effective potential $e [\va]$. In the Hamiltonian approach the effective potential $e [\va]$ of a spatial background field $\va$ is given by the minimum of the 
energy density $\vev{H}/V$ calculated under the constraint $\langle \vA \rangle = \va$. This minimal property of the effective potential calls for a 
variational calculation.

\section{Hamilton approach in background gauge}\label{sectionII}

In the presence of an external constant background field $\va$ the Hamiltonian approach can be most conveniently formulated in the 
background gauge
\be
\label{4}
\left[ \vd, \vA \right] = 0 \, , \quad \vd = \vec{\partial} + \va \,,
\ee
where all fields are taken in the adjoint representation. This gauge allows for an explicit resolution of Gauss' law, so that the gauge
fixed Hamiltonian can be obtained in explicit form \cite{Reinhardt:2013iia}.
 
We are interested here in the energy density in the state $\psi_a [A]$ minimizing
$
\vev{ H }_a \coloneq \bra{\psi_a } H \ket{\psi_a }
$
under the constraint $\langle \vA \rangle_a = \va$. For this purpose we perform a variational calculation with the trial
wave functional 
\be
\label{10}
\psi_a [A]  =  J [A]^{- 1/2} \tilde{\psi} [A - a]\,,\quad
\tilde{\psi} [A]  = \cN \e^{  - \frac{1}{2} \int A \omega A }\, ,
\ee
where 
$
J [A] = \Det (- \vD \cdot \vd) \, , \, \vD = \vec{\partial} + \vA \,
$
is the Faddeev--Popov determinant. This ansatz
 already fulfills the constraint $\langle \vA \rangle_a = \va$ and reduces 
for $\va = 0$ to the trial wave functional used
in Coulomb gauge \cite{FeuRei04}. Furthermore the variation kernel $\omega$ has the meaning of the gluon energy.
Proceeding as in the variational approach in Coulomb gauge \cite{FeuRei04}, from 
$\langle H \rangle_a \to \text{min}$ one derives a set of coupled equations for the gluon and ghost propagators.
Using the same approximations as in ref.~\cite{HefReiCam12} in Coulomb gauge, i.e. 
restricting to two loops in the energy, while neglecting the so-called Coulomb term
 and also the tadpole arising from the non-Abelian part of the magnetic energy, one finds from the
minimization of $\langle H \rangle_a$ the following gap equation 
\be
\label{12}
\omega^2 =   - {\vd} \cdot {\vd}  + \chi^2\,,
\ee
where  $\chi$ is the ghost loop (referred to as ``curvature'') see ref.~\cite{Reinhardt:2013iia}. 
Lattice calculations \cite{BurQuaRei09} of the gluon propagator in Coulomb gauge show that the gluon energy can be nicely fitted by Gribov's formula \cite{Gribov78}
\be
\label{22}
\omega ( | \vp|) = \sqrt{\vp^2 + M^4/\vp^2} \, .
\ee
A full self-consistent solution of the gap equation (\ref{12}) and the ghost DSE reveals that $\omega (p)$ contains in addition sub-leading UV-logs,
which on the lattice are found to be small. 
\section{The effective potential}
\begin{figure}[t]
\centering
\parbox[t]{.43\linewidth}{
\includegraphics[width=\linewidth]{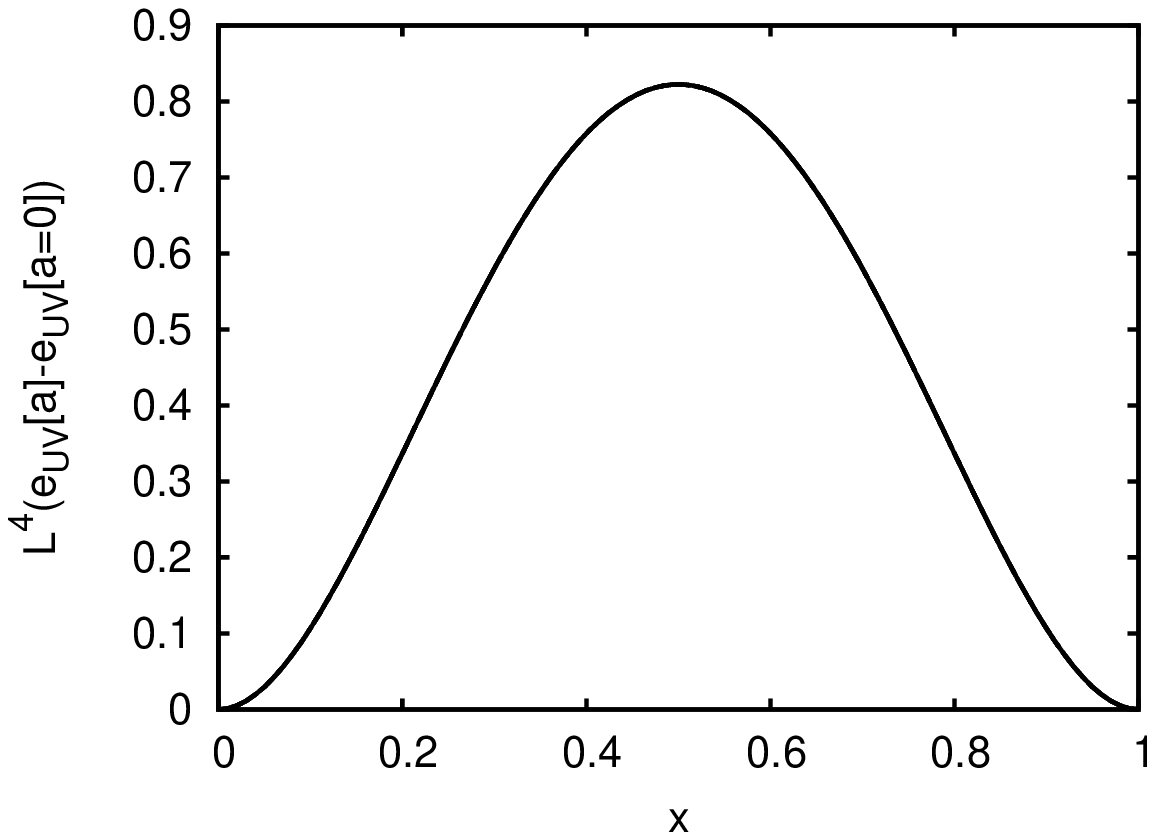}
\caption{The Weiss potential $e_\text{UV}$ (\protect\ref{27}).}
\label{fig1-1}
}
\hfil
\centering
\parbox[t]{.43\linewidth}{
\includegraphics[width=\linewidth]{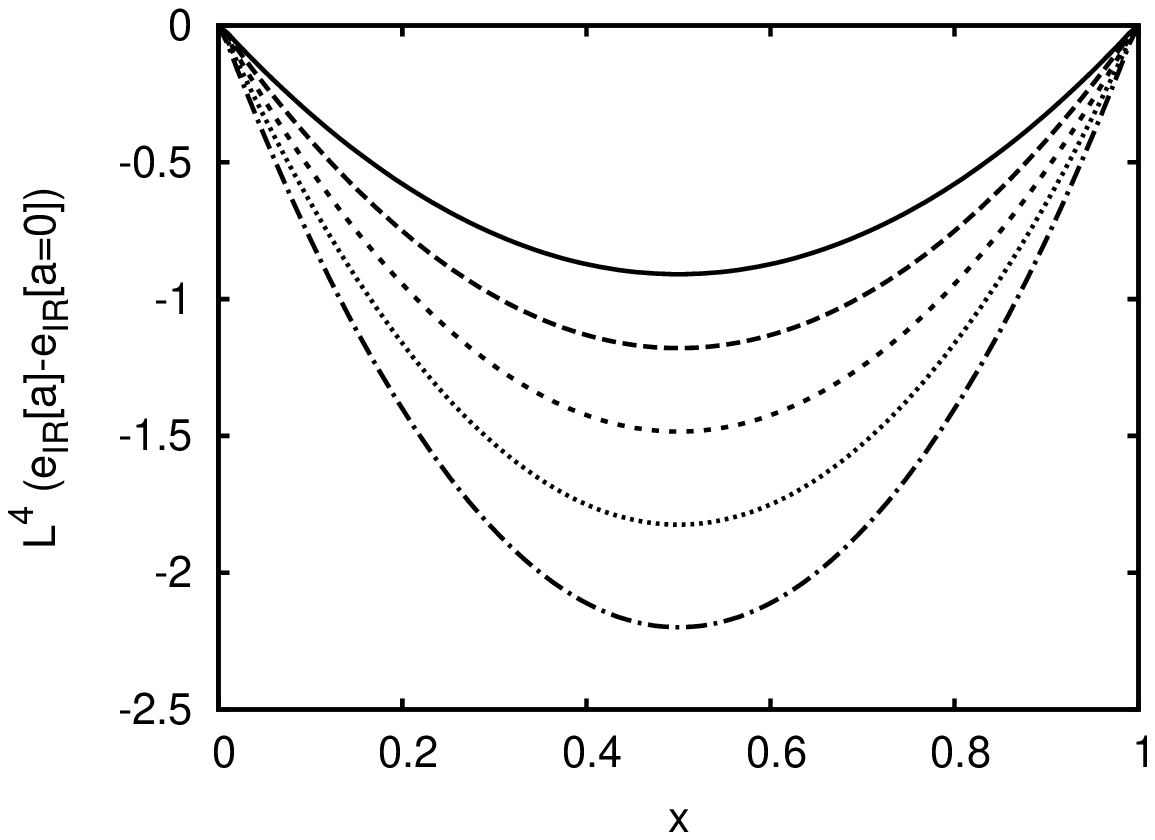}
\caption{The infrared potential $e_\text{IR}$ (\protect\ref{28}).}
\label{fig1-2}
}
\end{figure}
The background gauge field enters the background gauge fixed Hamiltonian only via the covariant 
derivative $\vd$ (\ref{4}) in the adjoint representation. 
It is therefore convenient  to go to the Cartan basis in which the generators of the 
Cartan subalgebra are diagonal. In the adjoint representation, their eigenvalues $\sigma_k$ form
the root vectors ${\sigma}  =(\sigma_1, \sigma_2, \ldots, \sigma_r)$, where $r$ is the rang of the group ($r = 1$ for SU$(2)$ and $r = 2$ 
for SU$(3)$). Compactifying the $3$-axis to a circle with circumference $L$ and choosing the background field along the compactified dimension $\va = a \ve_3$
the eigenvalues of $- \ii \vd = -\ii (\vec{\partial} + \va )$ read
\be
\label{24}
\vp^\sigma = \vp_\perp + \lk p_n  \sigma \cdot  a \rk \ve_3 \, , \quad p_n = 2 \pi n/L \, ,
\ee
where $\vp_\perp$ is the projection of $\vp$ into the $1$-$2$-plane and $\sigma \cdot  a = \sum_{k=1}^r \sigma_k a_k$, with the components $a_k$
of the gauge field along the generators $H_k$ of the Cartan algebra. If $t_a$ ($a=1,\, \ldots N^2-1$) denotes the generators of the gauge group in the usual representation we have $H_1 = t_3$ for SU($2$) and for SU($3$) in addition $H_2 =t_8$.
\begin{figure}[t]
\centering
\parbox[t]{.43\linewidth}{
\includegraphics[width=\linewidth]{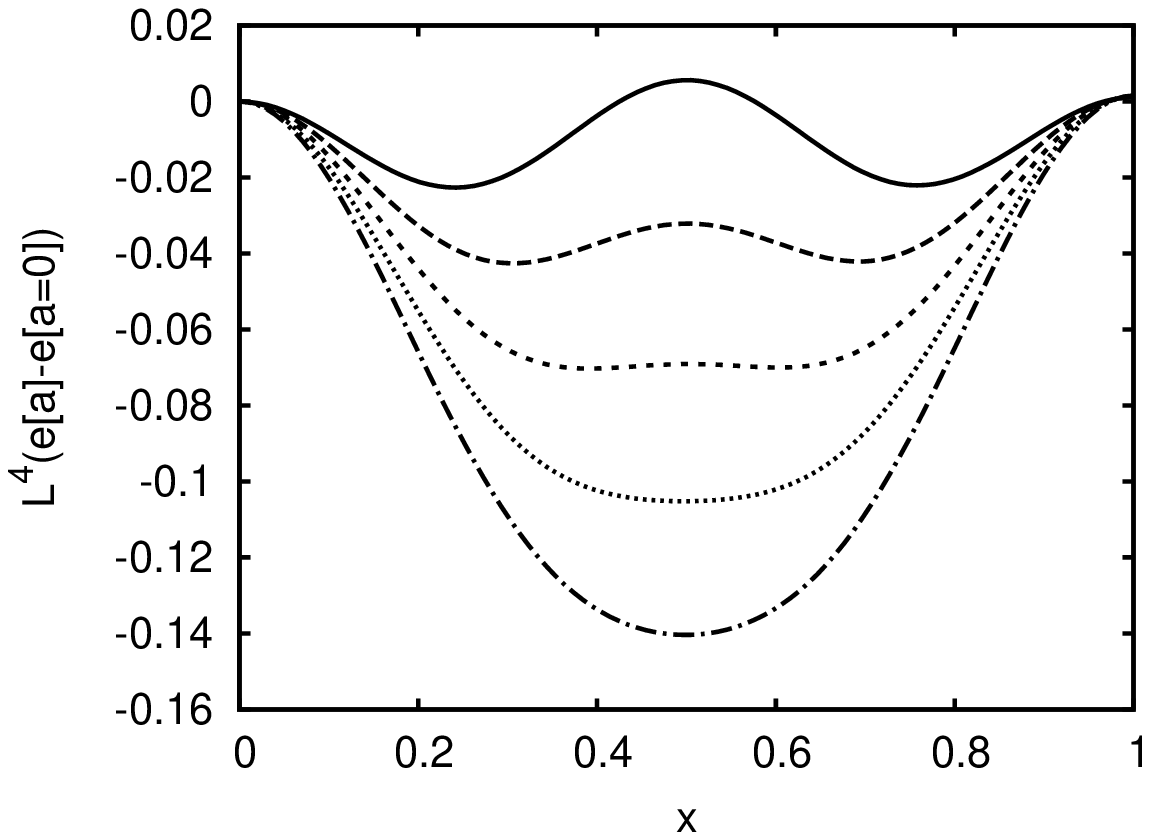}
\caption{The full effective potential for SU$(2)$ for different temperatures $L^{-1}$.}
\label{fig3}
}
\hfil
\centering
\parbox[t]{.43\linewidth}{
\includegraphics[width=\linewidth]{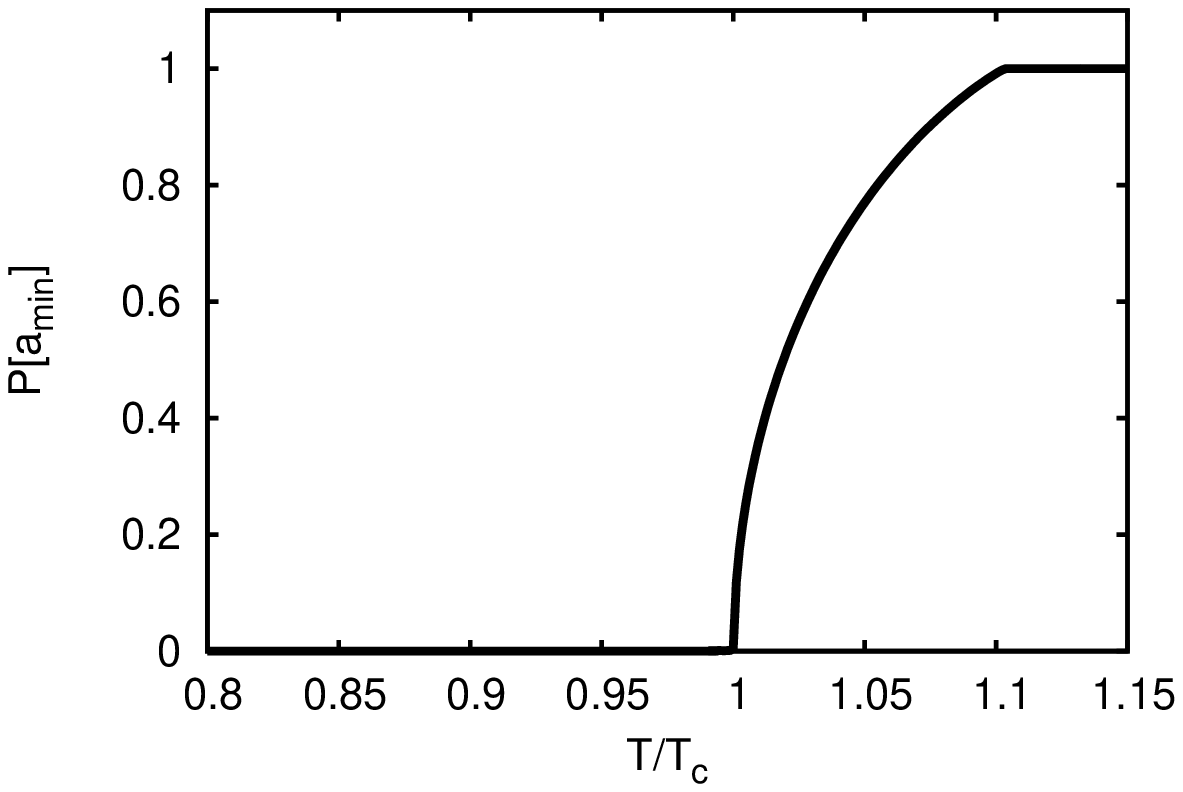}
\caption{The Polyakov loop $P[a^\text{min}]$ calculated at the minimum $a = a^\text{min}$ of the effective potential for SU($2$) as a function of $T/T_c$.}
\label{fig4}
}
\end{figure}
The effective potential is then obtained as \cite{Reinhardt:2012qe, Reinhardt:2013iia}
\begin{align}
\label{352-18}
e (a, L) = \sum\limits_\sigma \frac{1}{L} \sum\limits^\infty_{n = - \infty} \int \frac{\dd^2 p_\perp}{(2 \pi)^2} (\omega (p^\sigma) - \chi
(p^\sigma) ) \, , \quad p^\sigma = | \vp^\sigma|\,,
\end{align}
where $\omega$ and $\chi$ are the gluon energy and the ghost loop at zero temperature 
in Coulomb gauge, 
which, however, have to be taken here at the momentum argument (\ref{24}) 
shifted by the background field.
This potential has the required periodicity 
\begin{align}
\label{358-20}
e (a, L) = e (a + 2 \pi \mu_k / L, L) \, ,
\end{align}
where $\mu_k$ denotes the co-weights of the gauge algebra, which are related to the 
center elements $z_k \in  \mathrm{Z}(N)$ of the gauge group by
$\exp (\ii 2 \pi \mu_k) = z_k$.
The expression (\ref{352-18}) for the effective potential is surprisingly simple and requires only
the knowledge of the gluon energy $\omega$  and the ghost loop $\chi$ in Coulomb gauge 
at zero temperature. 

If one ignores the ghost loop
$\chi (p) = 0$ the potential (\ref{352-18}) becomes the energy density of a non-interacting 
Bose gas with single-particle energy $\omega (p)$, living, however, on the spatial manifold 
$\bR^2 \times S^1$. With $\chi (p) = 0$ and replacing the gluon energy $\omega (p)$ (\ref{22}) 
by its ultraviolet part $\omega_\mathrm{UV} (p) = | \vp|$ one obtains from (\ref{352-18}) precisely the Weiss potential
\cite{Weiss:1980rj}
\be
\label{27}
e_\mathrm{UV} (a, L) = \frac{4}{3} \frac{\pi^2}{L^4}  x ^2 \lk x- 1 \rk^2 \,,\quad x \equiv a L/(2 \pi)\,,\quad 0 \leq x \leq 1
\ee
corresponding to the deconfined phase. If on the other hand one chooses the infrared form of the 
gluon energy (\ref{22}) $\omega_\mathrm{IR} (p) = {M^2}/{|\vp|}$ one obtains the potential
\be
\label{28}
e_\mathrm{IR} (a, L) = 2 \frac{M^2}{L^2} \left(  x ^2 - x \right) \,,\quad x \equiv a L/(2 \pi)\,,\quad 0 \leq x \leq 1\,,
\ee
which is shown in figure \ref{fig1-2}, whose minimum occurs at the center symmetric configuration, 
which yields a vanishing Polyakov loop corresponding to the confined phase. Obviously, the 
deconfinement phase transition results from the interplay between the confining IR-potential and the deconfining UV-potentials. Choosing $\omega (p) = \omega_\mathrm{IR} (p) + \omega_\mathrm{UV} (p)$,
which can be considered as an approximation to the Gribov formula (\ref{22}), one has to add
the UV- and IR-potentials, given by eqs.~(\ref{27}) and (\ref{28}), respectively,
 and finds a phase transition at a critical temperature $T_c = \sqrt{3} M / \pi$. With the Gribov mass
$M = 880$ MeV this gives a critical temperature of $T_c \approx 485$ MeV, which is much too high. 
One can show analytically, see ref.~\cite{Reinhardt:2013iia}, that the neglect of the
ghost loop $\chi (p) = 0$ shifts the critical temperature to higher values. If one uses eq. 
(\ref{22}) for $\omega (p)$ and includes the ghost loop 
one finds the effective potential shown in fig.~\ref{fig3}, which gives a 
transition temperature $T_c \approx 269$ MeV for SU$(2)$, which is in the right ball park.
\begin{figure}[t]
\parbox[t]{.43\linewidth}{
\includegraphics[width=\linewidth]{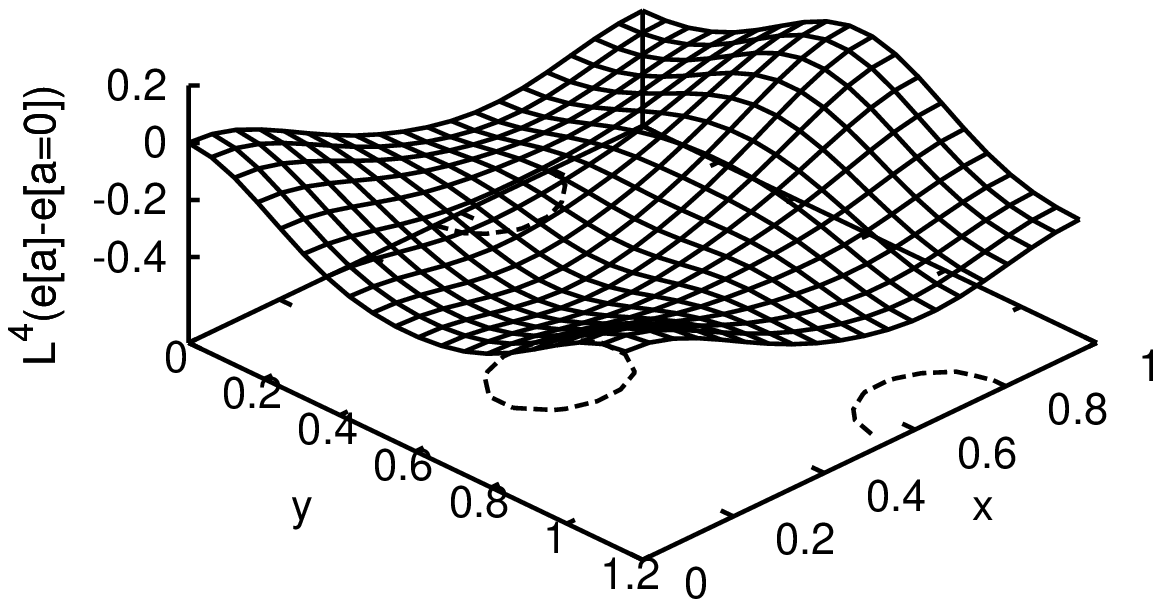}}
\hfil
\parbox[t]{.43\linewidth}{
\includegraphics[width=\linewidth]{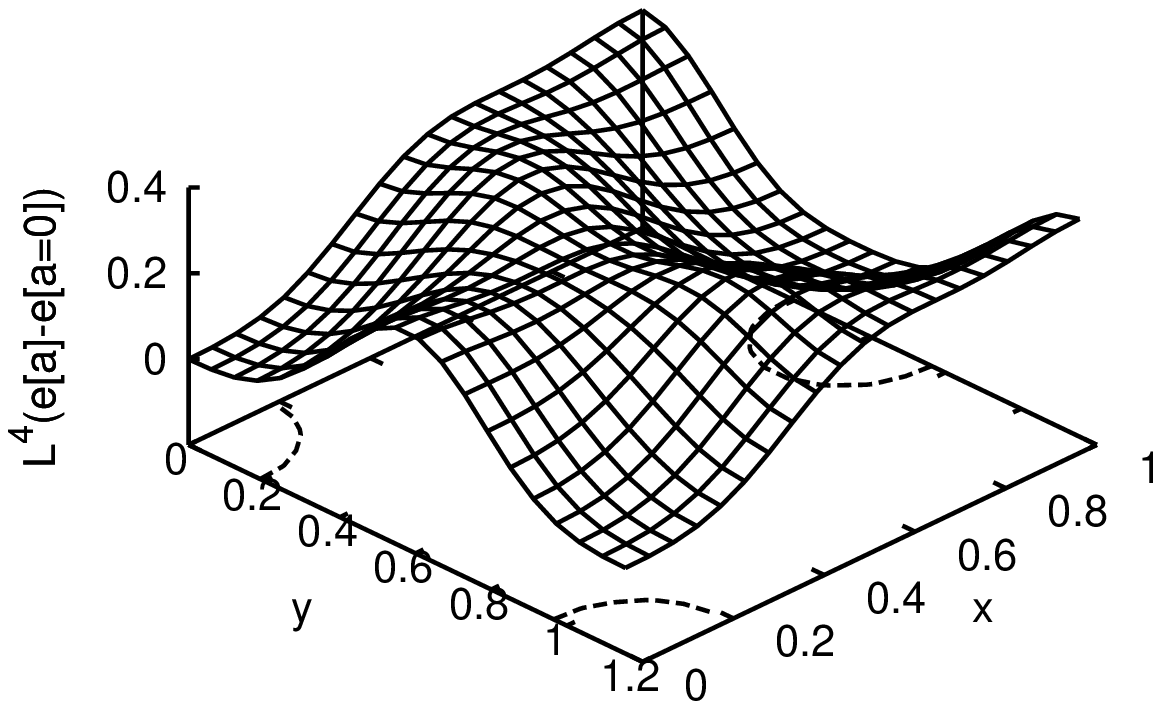}}
\caption{SU$(3)$ effective potential below (left panel) and above (right panel) $T_c$ as functions of $x=a_3 L/(2\pi)$ and $y=a_8 L/(2\pi)$.}
\label{fig-11} 
\end{figure}
The Polyakov loop $P[a^\text{min}]$ calculated from the minimum $a^\text{min}$ of the effective potential $e(a,L)$ is plotted in fig.~\ref{fig4}.

\begin{figure}[t]
\centering
\parbox[t]{.43\linewidth}{
\includegraphics[width=\linewidth]{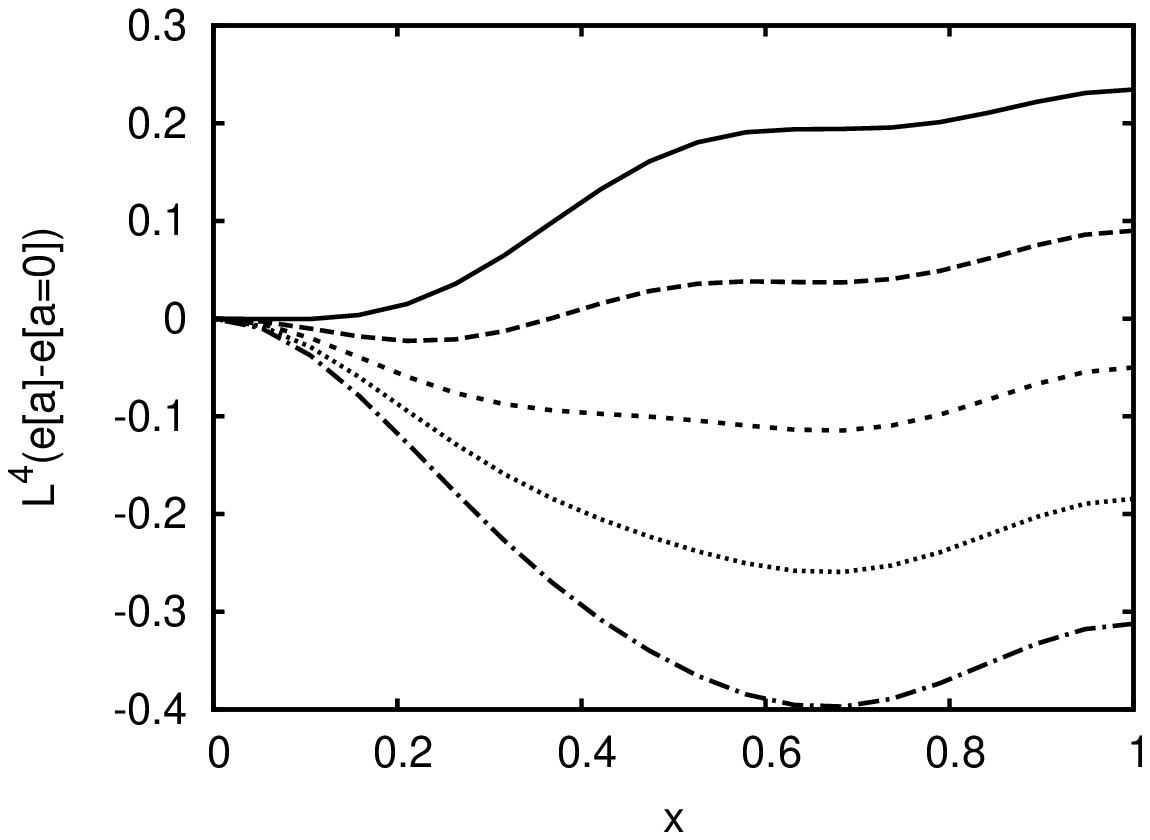}
\caption{SU($3$) effective potential, cut at $a_8 = 0$, for different temperatures $L^{-1}$.}
\label{fig-12a}  
}
\hfil
\centering
\parbox[t]{.43\linewidth}{
\includegraphics[width=\linewidth]{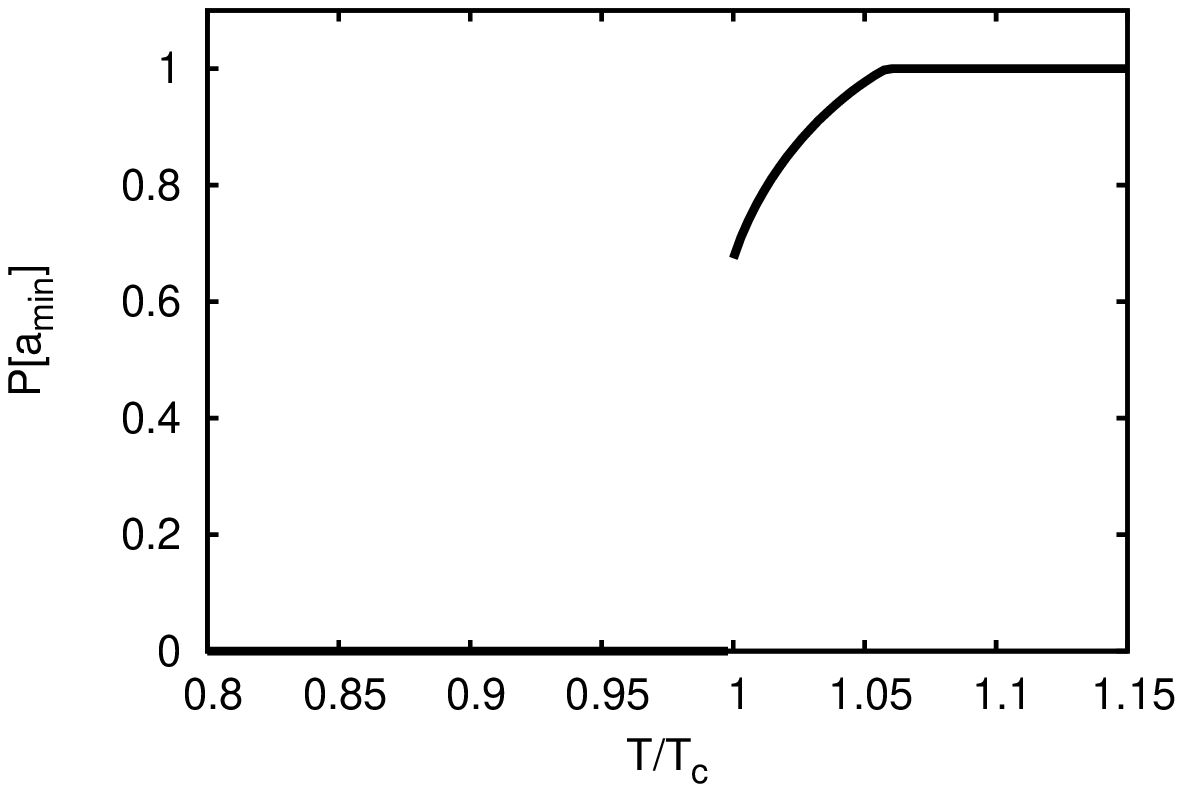}
\caption{The Polyakov loop $P[a^\text{min}]$ calculated at the minimum $a = a^\text{min}$ of the effective potential for SU($3$) as a function of $T/T_c$.}
\label{fig-12}  
}
\end{figure}

The effective potential for the gauge group  SU$(3)$ can be reduced to that of the SU$(2)$ group 
by noticing that the SU$(3)$ algebra consist of three SU$(2)$ subalgebras characterized by the 
three positive roots $\vsigma = (1, 0), \, \left( \frac{1}{2}, \frac{1}{2\sqrt{3}} \right), \, \left( \frac{1}{2}, -\frac{1}{2 \sqrt{3} }\right)$
resulting in 
\begin{align}
\label{418-25}
e_{\text{SU}(3)} (a) = \sli_{\sigma > 0} e_{\text{SU}(2)}[\sigma] (a) \, .
\end{align}
The effective potential for SU$(3)$ is shown in fig.~\ref{fig-11} as a function of $a_3$, $a_8$. 
As one notices, above and below $T_c$ the minima of the 
potential occur in both cases for $a_8 = 0$. Cutting the $2$-dimensional surfaces at $a_8 = 0$ one finds the effective
potential shown in fig.~\ref{fig-12a}. This shows a first order phase transition, which occurs at a critical temperature of $T_c = 283$ MeV.
The first order nature of the SU($3$) phase transition is also seen in fig.~\ref{fig-12} where the Polyakov loop $P[a^\text{min}]$ is shown.

\section{Conclusions}
\label{sec4}
In my talk I have shown that the effective 
potential of the Polyakov loop can be obtained from the zero-temperature energy density by compactifying one spatial dimension. In this approach the deconfinement phase transition is entirely determined by the zero-temperature propagators, which are defined as 
vacuum expectation values. Consequently, the finite-temperature behavior of the theory and, in particular, the dynamics of the deconfinement
phase transition must be fully encoded in the vacuum wave functional. The calculated effective potential yields also the correct order of the deconfinement
phase transition for SU$(2)$ and SU$(3)$. Presently the Hamiltonian approach in Coulomb gauge is extended to full QCD~\cite{Pak:2011wu,Pak:2013uba}. After extending the approach to full QCD we plan to consider the influence of an external magnetic field and to study the phase diagram at finite baryon density.

\end{document}